\documentstyle[preprint,aps]{revtex}

\begin{document}
\draft
\title{Majoron emitting neutrinoless double beta decay in the 
electroweak 
chiral gauge extensions}
\author{F. Pisano}
\address{Departamento de F\'\i sica, Universidade Federal do Paran\'a, 
81531-990, Curitiba, PR, Brazil}
\author{S. Shelly Sharma}
\address{Departamento de F\'\i sica, Universidade Estadual de Londrina,
86051-970, Londrina, PR, Brazil}
\date{\today}
\maketitle
\begin{abstract}
Fundamental mechanisms for Majoron emitting neutrinoless double beta
decay in SU(3)$_C\otimes G_W\otimes $ U(1) models, for electroweak
flavor chiral extensions, $G_W=$ SU(3)$_L$ and SU(4)$_L$ are pointed
out.  Both kinds of known Majoron emitting processes, charged Majoron
emitting where the massless Nambu-Goldstone boson itself carries
lepton charge, $L=-2$, and the ordinary Majoron emitting where the
boson has a small mass are found possible.
\end{abstract}
\bigskip
\pacs{PACS numbers: 11.15.Ex, 12.60.Fr, 14.80.Cp}




\bigskip

\newpage

Although two neutrino double beta decay \cite{Mayer35}, which can occur as a
second order weak interaction process within the minimal electroweak
standard model \cite{Glashow61}, has been observed experimentally \cite
{Elliot87}, there is no conclusive evidence for the neutrinoless modes \cite
{Furry39}. The search for $\beta \beta _{0\nu }$ and $\beta \beta _{\text{M}%
}$ decay where two outgoing electrons are accompanied by a Nambu Goldstone
boson called the Majoron, is motivated by the promise that the observation
of these processes may hold the key to physics beyond the standard model.
Besides the minimal standard model, there is no particular reason for
defending the concept of massless neutrinos, thus discriminating between
these and other massive fermions. If the neutrinos are massive they may be
either Dirac or Majorana fermions. In the minimal standard model the
neutrinos are strictly massless Weyl fermions since a right handed neutrino
world is absent. Dirac mass is not possible for neutrinos in such a scenario
as the left and right-handed neutrinos represent disconnected independent
degrees of freedom. On the other hand if the lepton number is an additive
exactly conserved charge, Majorana mass is forbidden. Any attempt to
generate neutrino masses has to transgress at least one of these two
assumptions. Concerning the second possibility, present experimental
evidence and the standard model are consistent with the absolute
conservation of three separate $L_l$ , $l=\{e,\mu ,\tau \}$ lepton numbers.
The search for neutrinoless double beta decay provides limits for $\Delta
L=2 $ violations for one type of leptons. The best laboratory limits
available to date are $t_{1/2}>5.6\times 10^{24}$ yrs 
with 90\% of confidence level
for $^{76}$Ge $\rightarrow $ $^{76}$Se $+$ $2e^{-}$ decay \cite{PDG96} and 
the recently reported $t_{1/2}>7.4\times 10^{24}$ yrs ($90\% $ C.L.) for the
same decay from the Heidelberg-Moscow experiment \cite{Gunt97}. If
lepton number violation of the $\Delta L=2$ type occurs, the neutrinos are
expected to have a nonvanishing Majorana mass. The $\beta \beta _{0\nu }$
and $\beta \beta _{\text{M}}$ decay being $\Delta L=2$ processes, the
observation of either would certainly imply new physics.

The neutrinoless decay mode accompanied by Majoron emission, might hold a
clue to a new fundamental interaction. The Majoron, a Nambu-Goldstone boson,
was originally associated with spontaneous breaking of a U(1) lepton number
symmetry \cite{Chik80}-\cite{Geor81} or as proposed recently \cite{Burg94},
it could be a Nambu-Goldstone boson for a symmetry distinct from that of the
lepton number, which carries the classical unbroken lepton number charge.
Theories containing such a Majoron are called charged Majoron models and the
broken symmetry could be gauged \cite{Caro93}.

While the standard model of nongravitational interactions is extremely
successful and consistent with the existing experimental data (some data not
easily accommodated in the model not being definitive as yet), it leaves
some fundamental questions unexplained. One of these questions is the family
replication problem or the flavor question. Why are there exactly three
families of leptons and quarks?

There has been a renewed interest, during the last few years, in the
prospect of enlarging the weak isospin factor $G_W$ in the gauge group

\[
G_0=\text{SU(3)}_C\otimes G_W\otimes \text{U(1)}_X, 
\]
so that explanations to some of the fundamental questions like the flavor
question, could be found. In fact, in the standard model \cite{Glashow61},
each family of fermions is anomaly free. Interestingly it is true for many
extensions of the standard model as well, including the popular grand
unified theories \cite{Geor74}. Therefore, in these models, there is no
restriction on the number of families on theoretical grounds. Nowhere in
physics the flavor question is replied \cite{Glashow83} except in $G_0$
models where each family is anomalous but unlike in standard 
model, different families are not exact
replicas of one another. The anomalies cancel out
when the number of families is an integer multiple of the number of color
degrees of freedom. The most economical $G_{0\text{ }}$group that admits
such a fermion representation is the weak isospin enlarged group for 
$G_W=\text{SU(3)}_L$ \cite{Pisano92}-\cite{Mont93}. Using the lightest leptons,
including right-handed neutrinos, as the particles which determine the
approximate $G_W$ symmetry and treating each family separately, $G_W=$SU(4)$%
_L$ is the highest symmetry group to be considered in the electroweak
sector. A model with the SU(4)$_L\otimes $ U(1) gauge symmetry in the lepton
sector was suggested some years ago by Voloshin \cite{Volo88}. This symmetry
in both quarks and leptons has been pointed out recently \cite{Foot94}.

Some of the novel features of $G_0$ gauge groups which are lost in the
standard model are: 1) leptons are treated democratically in each of the
three families. Each individual family possesses nonvanishing anomalies and
only with a matching of the number of families with the number of quark
colors does the overall anomaly vanish. This novel mechanism of anomaly
cancellation solves the flavor question \cite{Pisano96}; 2) the electroweak
mixing angle is constrained by the Landau pole of the model; 3) there is a
connection between the neutrino mass terms and the electromagnetic gauge
invariance \cite{Ozer94}; 4) the new physics is guaranteed to be below a few
TeV, well within the reach of the next generation colliders \cite{Ng94}; 5)
just as there is a natural answer to the family replication question there
are some indications as to why the top quark is so heavy. Possibly these
models offer the right approach to the question of fermion mass generation.
The third family could be the door to new physics; 6) with suitable
fermionic representation content in $G_W=$ SU(3)$_L$ or SU(4)$_L$ it is
possible to realize the Voloshin mechanism \cite{Volo88} that allows
neutrinos to have magnetic moment even in the massless limit\cite{Barb89}.

Let us consider the specific model \cite{Pisano92} with the representation
content as given below. The leptons transform as

\begin{equation}
\Psi _{aL}\equiv \left( 
\begin{array}{c}
\nu _a \\ 
l_a \\ 
l_a^c
\end{array}
\right) _L\sim \left( {\bf 1},{\bf 3}_{{\bf L}},0\right)  \label{eq1}
\end{equation}
for the three families $a=\{e,\mu ,\tau \}$. In the leptonic sector we have
the Yukawa interactions

\begin{equation}
L_{lS}=-\frac 12\sum\limits_{a,b}G_{ab}\overline{(\Psi _{iaL})^c}\smallskip\
\Psi _{jbL}\smallskip\ S_{ij}+\text{H.c}.  \label{eq2}
\end{equation}
with $\Psi ^c=C\stackrel{}{\overline{\Psi }^T,}C$ being the charge
conjugation matrix, $i,j$ SU(3) indices, and $S_{ij}$ elements of the
symmetric sextet of scalar fields,

\begin{equation}
S=\left( 
\begin{array}{ccc}
\sigma _1^0 & h_2^{+} & h_1^{-} \\ 
h_2^{+} & H_1^{++} & \sigma _2^0 \\ 
h_1^{-} & \sigma _2^0 & H_2^{--}
\end{array}
\right) \sim ({\bf 1},{\bf 6}_{{\bf s}}^{*},0),  \label{eq3}
\end{equation}
which contains the Gelmini-Roncadelli \cite{Gelm81} SU(2) triplet, the SU(2)
doublet of the electroweak standard model \cite{Glashow61} and the doubly
charged singlet that appears in the model of Babu \cite{Babu88}. Explicitly
these interactions are

\begin{eqnarray}
L_{lS} &=&-\frac 12\sum_{a,b}G_{ab}[\ \overline{\nu _{aR}^c}\ \nu _{bL}\
\sigma _1^0+\overline{l_{aR}^c}\ l_{bL}\ H_1^{++}+\overline{l_{aR}}\
l_{bL}^c\ H_2^{--}  \nonumber \\
&&+\left( \overline{\nu _{aR}^c}\smallskip\ l_{bL}+\overline{l_{aR}^c}%
\smallskip\ \nu _{bL}\right) h_2^{+}+\left( \overline{\nu _{aR}^c}\smallskip%
\ l_{bL}^c+\overline{l_{aR}}\smallskip\ \nu _{bL}\right) h_1^{-}  \label{eq4}
\\
&&+\left( \overline{l_{aR}^c}\smallskip\ l_{bL}^c+\overline{l_{aR}}\smallskip%
\ l_{bL}\right) \sigma _2^0\smallskip\ ]+\text{H.c.}  \nonumber
\end{eqnarray}
and if we impose the condition $\left\langle \sigma _1^0\right\rangle =0$ ,
then the neutrinos remain massless, at least at the tree level.

Interesting mechanisms for neutrinoless double-beta decay within $G_0$
models were considered in Ref. \cite{Plei93}. There it was noted that the
neutrinoless double-beta decay has contributions that do not depend
explicitly on neutrino mass. In comparison with most extensions of the
electroweak model, neutrinoless double-beta decay mechanism requires less
neutrino mass in $G_0$ models. However, the smallness of neutrino mass has
no relation with the bad high energy behavior of processes such as $%
W^{-}W^{-}\rightarrow e^{-}e^{-},$ since in these models the doubly charged
gauge boson $U^{--}$ which is a component of the dilepton gauge boson
doublet ($U^{--},V^{-}$), cancels out the divergent part of such a process.

Let us define the additively conserved leptobarion number as

\begin{equation}
F=L+B  \label{eqf1}
\end{equation}
where $L=\sum\limits_aL_a,$ $a=\{e,\mu ,\tau \}$ is the total lepton number
and $B$ is the baryon number. For any lepton $l$,

\begin{equation}
F(l)=F(\nu _l)=+1  \label{eqf2}
\end{equation}
and in order to make $F$ a conserved quantum number in the Yukawa
interactions we assign to the scalar fields of the sextet the values

\begin{equation}
-F(H_2^{--})=F(H_1^{++})=F(h_2^{+})=F(\sigma _1^0)=-2  \label{eqf3}
\end{equation}
The conservation of the leptobarion number $F=L+B$ forbids the existence of
massive neutrinos and the neutrinoless double beta decay through mechanisms
discussed in Ref. \cite{Plei93}. For neutrinoless double-beta decay to
occur, the $F$ symmetry must be broken. A spontaneous breaking of $F$
symmetry, in this case, implies a Majoron-Goldstone-like boson because $%
\sigma_1^0,$ $h_2^{+}$ and $H_1^{++}$ constitute a triplet under SU(2). The 
vertex interactions $\overline {\nu_{eR}^c}\ \nu _{eL}^{}\ \sigma _1^0$ and $%
\overline{e_R}\smallskip\ e_L \smallskip\ \sigma _2^0$ in the Yukawa
couplings of Eq. (\ref{eq4}) give rise to processes for double-beta decay
with Majoron emission as shown in Figs. 1 and 2. In particular, the process
of Fig. 2 occurs only if direct electron-Majoron couplings are possible.

Similar kind of Majoron emitting double-beta decay mechanisms arise in the $%
G_{\text{0}}$ model with $G_W=$ SU(4)$_L$. If right-handed neutrinos are
introduced it is an interesting possibility to have $\nu $, $e,$ $\nu ^c$
and $e^c$ in the same multiplet of SU(4) $\otimes $ U(1) gauge electroweak
group. If each family of fermions is treated separately, using the lightest
leptons as the particles which determine the approximate symmetry, SU(4) is
the highest symmetry group to be considered in the electroweak sector. Now
leptons transform under $G_0$ as

\begin{equation}
\Psi _{aL}\equiv \left( 
\begin{array}{c}
\nu _a \\ 
l_a \\ 
\nu _a^c \\ 
l_a^c
\end{array}
\right) _L\sim \left( {\bf 1},{\bf 4}_{{\bf L}},0\right)  \label{eqsu(4)1}
\end{equation}
with the Yukawa interactions

\begin{equation}
L_{lH}=-\frac 12\sum\limits_{a,b}G_{ab}\overline{(\Psi _{iaL})^c}\smallskip\
\Psi _{jbL}\smallskip\ H_{ij}+\text{H.c}.  \label{eqsu(4)2}
\end{equation}
where $H_{ij}$ are the elements of the symmetric decuplet of scalar fields

\begin{equation}
H=\left( 
\begin{array}{cccc}
H_1^0 & H_1^{+} & H_2^0 & H_2^{-} \\ 
H_1^{+} & H_1^{++} & H_3^{+} & H_3^0 \\ 
H_2^0 & H_3^{+} & H_4^0 & H_4^{-} \\ 
H_2^{-} & H_3^0 & H_4^{-} & H_2^{--}
\end{array}
\right) \sim ({\bf 1},{\bf 10}_{{\bf s}}^{*},0),  \label{eqsu(4)3}
\end{equation}
which contains the symmetric sextet $S$ of Eq. (\ref{eq3}). Charged leptons
get a mass if $\left\langle H_3^0\right\rangle \neq 0$ and with this vacuum
structure neutrinos remain massless at tree level. All possible couplings of
leptonic bilinears with scalar fields are contained in the following
expansion of $L_{lH}$

\begin{eqnarray}
L_{lH} &=&-\frac 12\sum\limits_{a,b}G_{ab}[\overline{\nu _{aR}^c}\ \nu
_{bL}\ H_1^0+\overline{l_{aR}^c}\ l_{bL}\ H_1^{++}+\overline{l_{aR}}\
l_{bL}^c\ H_2^{--}+\overline{\nu _{aR}}\ \nu _{bL}^c\ H_4^0  \nonumber
 \\
&&+(\overline{\nu _{aR}^c}\ l_{bL}+\ \overline{l_{aR}^c}\ \nu _{bL}\ )%
\smallskip\ H_1^{+}+(\overline{\nu _{aR}^c}\ \nu _{bL}^c+\ \overline{\nu
_{aR}}\ \nu _{bL}\ )\smallskip\ H_2^0  \nonumber \\
&&+(\overline{\nu _{aR}^c}\ l_{bL}^c+\ \overline{l_{aR}}\ \nu _{bL})\
H_2^{-}+(\overline{l_{aR}^c}\ l_{bL}^c+\overline{l_{aR}}\ l_{bL})\ H_3^0 
\nonumber \\
&&+(\overline{l_{aR}^c}\ \nu _{bL}^c+\overline{\nu _{aR}}\ l_{bL})\ H_3^{+}+(%
\overline{\nu _{aR}}\ l_{bL}^c+\overline{l_{aR}}\ \nu _{bL}^c)\ H_4^{-}]+%
\text{H.c.}  \label{eqsu(4)4}
\end{eqnarray}
The terms $\overline{\nu _{aR}^c}\ \nu _{bL}\ H_1^0$ for example are the
analog of $\overline{\nu _{aR}^c}\ \nu _{bL}\ \sigma _1^0$ of Eq. (\ref{eq4}%
). Similarly we can identify $(\overline{l_{aR}^c}\ l_{bL}^c+\overline{l_{aR}%
}\ l_{bL})H_3^0$ as being similar to the terms involving coupling with the
scalar field $\sigma _2^0$ in $L_{lS}.$

The total decay rate for double beta decay processes involves the model
dependent coupling constants and mixing parameters, the relevant nuclear
matrix elements and a factor determined by the phase space for a particular
process. Following the work of Vogel and Zirnbauer\cite{Voge86}, the
proton-neutron quasiparticle random phase approximation (pn-QRPA) has turned
out to be the most popular model and good estimates for double beta
transition matrix elements have been obtained in several works. However 
the calculated nuclear matrix elements reported in different
theoretical works show a large variance.
Experimentally observed $\beta \beta _{2\nu }$ decay rates serve as a
constraint on the nuclear structure calculations. 
The calculated 
nuclear matrix elements for $\beta \beta_{2 \nu}$ as well as 
$\beta \beta_{0 \nu}$ transitions
are seen to be \cite{Krmp94}  extremely sensitive to the proton-neutron 
interaction parameters. In general the Gamow-Teller matrix elements for 
$\beta \beta_{0 \nu}$ decay, calculated within the pn-QRPA framework but 
using different
calculation schemes and interactions differ from each other by a factor of
 upto three.
In a shell model study of double beta decay of the nuclei $ ^{76}Ge$,
 $ ^{82}Se$   
and $ ^{136}Xe$ {\cite{Caur96}}, the calculated nuclear matrix elements 
for $\beta \beta_{0 \nu}$
are found to be in general much
smaller than those reported for the same nuclei by using pn-QRPA.
 Barbero et al \cite{Barb96}
have calculated the nuclear matrix elements for charged Majoron emission,
using the quasiparticle random phase approximation method of Refs. \cite
{Hirs90,Krmp94} for the model given in \cite{Burg94}. Some of the nuclear
matrix elements contributing to the charged Majoron emission process are
distinct from the ones involved in $\beta \beta _{2\nu }$ and $\beta \beta
_{0\nu }$ decay processes.  
Nuclear matrix elements for various
types of Majoron emitting processes have also been reported 
in Ref{\cite{Hirs96}}. 
Experimental half life for a given type of double beta decay 
process or a limit thereof can, together
with theoretically calculated nuclear matrix elements, be used to obtain 
bounds on the model parameters. The bounds obtained in this fashion obviously
carry the uncertainities of the nuclear matrix elements
used in the calculation.

Evaluation of Majoron-emitting decay amplitude in Fig. 1 requires a
knowledge of the neutrino-Majoron coupling contained in Eq. (\ref{eq2}). For
neutrino masses that are much smaller than the Fermi momentum and neglecting
the final-state lepton energies and momenta, the expression for amplitude
simplifies to

\begin{equation}
{\cal A}(\beta \beta _{\text{OM}})\simeq -4\sqrt{2}\sum%
\limits_{ij}V_{ei}V_{ej}G_{ij}^R\int \frac{d^4p}{(2\pi )^4}\frac{W_\alpha ^{%
\smallskip\ \alpha }}{p^2+i\varepsilon }  \label{ampeq1}
\end{equation}
where the electron flavor rows of the associated leptonic Kobayashi-Maskawa
mixing matrix for the weak charged-current interactaions are denoted by $%
V_{ei}$ and $W_\alpha ^{\smallskip\ \alpha }$ are the form factors. 
We may define the effective Yukawa coupling of the Majoron to 
the electron neutrino to be  
$ g_{\text {eff}} = \sum \limits_{ij}V_{ei}V_{ej}G_{ij}^R$. For
charged Majoron case, taking into account different values of the spectral
index, $n_{\text{CM}}=3,$ $n_{\text{OM}}=1$, we have in the rest frame of
the nucleus

\begin{equation}
{\cal A}(\beta \beta _{\text{CM}})\simeq 8\sqrt{2}\sum%
\limits_{ij}V_{ei}V_{ej}G_{ij}^R\int \frac{d^4p}{(2\pi )^4}\left[ \frac{{\bf %
p}^2(w_5+w_6)}{(p^2-m_i^2+i\varepsilon )(p^2-m_j^2+i\varepsilon )}\right]
\label{ampeq2}
\end{equation}
with the parameters $w_5$, and $w_6$ given by

\[
w_5=\frac{p_i(W_{0i}-W_{i0})}{2\smallskip\ {\bf p}^2}\text{ and }w_6=\frac{%
\varepsilon _{ijk\smallskip\ }p_i\smallskip\ W_{jk}}{2\smallskip\ {\bf p}^2}%
\smallskip\ . 
\]
In the Eqs. (\ref{ampeq1}) and (\ref{ampeq2}), $G_{ij}^R$ denote the
coupling strength associated with the right-handed chiral projector in the
Yukawa couplings. We can immediately compare the amplitudes ${\cal A}(\beta
\beta _{\text{OM}})$ and ${\cal A}(\beta \beta _{\text{CM}})$ with the
amplitude for pure neutrinoless double beta decay, 
\begin{equation}
{\cal A}(\beta \beta _{0\nu })\simeq 8\sqrt{2}\pi \sum\limits_iV_{ei}^2%
\smallskip\ m_i\int \frac{d^4p}{(2\pi )^4}\left[ \frac{W_\alpha ^{\smallskip%
\ \alpha }}{(p^2-m_i^2+i\varepsilon )}\right]  \label{ampeq3}
\end{equation}
which vanishes in the absence of a direct Majorana mass for the electron
neutrino.

As pointed out earlier,in the $G_0$ gauge extensions scheme the following $J$
Majoron emitter process is possible 
\begin{equation}
(Z,A)\rightarrow (Z+2,A)+2e^{-}+J  \label{quinze}
\end{equation}
wherein no real neutrinos are produced. 
We may observe that due to increased
degrees of freedom the $G_0$ gauge extensions scheme admits 
the possibility of a contribution to $\cal{A}
(\beta \beta _{\text M} )$ 
of the type coming from an Ordinary Majoron
as well as a contribution where the Majoron is a charged Majoron.  We
recall that if we impose the condition, $ <\sigma_1^0>=0$, then neutrinos
remain massless at the tree level and $\beta \beta_{0 \nu}$ is 
forbidden by the 
lepton number conservation.  The scalar $\sigma_2^0$ gets a VEV and is
responsible for lepton mass generation.  As the emitted massless Nambu
Goldstone Boson carries a leptobarionic charge $F=-2$, the majoron emitting
 decay
of the 'charged majoron' type can occur.  
For this kind of process the
bound on Majoron-neutrino effective coupling can be calculated by using the 
reported experimental limit of $T_{\frac{1}{2}} > 5.85 X10^{21} $ yr 
on charged majoron emitting decay from the 
Heidelberg-Moscow experiment \cite{Gunt96}. In case the nuclear matrix 
elements and phase space from reference \cite{Hirs96} are used the calculated 
bound is $g_{\text{eff}} <\,0.18\, (90\% C.L. ) $.
 
Now if $<\sigma_1^0>\, \neq 0$, we have a spontaneous breaking of the $F$
symmetry, implying a Majoron Goldstone like Boson
 since $\sigma_1^0$ carries $F=-2$.  As mentioned 
 in Ref. \cite{Plei93}, from the experimental 
value of the mass of $\rho$, one
obtains $<\sigma_1^0>\, < 10$ MeV. The resulting Majoron emitting double 
beta decay process
is very similar to that in Gelmini-Roncadelli model\cite{Gelm81} with the 
important difference that 
the scalar $\sigma_1^0$ belongs to a
symmetric sextet of scalar fields .  In $G_0$ model only a VEV in the Fermi
scale contributes to the mass of Z, as such there is no 
conflict with the
measured width of Z standard boson. 
Majoron neutrino coupling strength for this kind 
of Majoron emitting process has
a bound of $g_{\text{eff}} < 2.3X10^{-4} $ now using the experimental limit
 of $T_{\frac{1}{2}}>7.91X10^{21}  $ yr from the
 Heidelberg-Moscow experiment \cite{Gunt96}  and the corresponding
 nuclear matrix element from Ref. \cite{Hirs96}.
These bounds 
serve to establish constraints on the $G_0$ models. In fact due to the specific
features of the $G_0$ models, the mass limits on $%
\sigma_1^0$ boson may be quite different from those for the boson
appearing in the triplet of Gelmini-Roncadelli \cite{Gelm81}.
The phenomenology of the $\sigma_1^0$ needs to be studied in more detail to 
make sure which one of the two possibilities mentioned above is more natural.

It is very difficult to distinguish
the Majoron emitting process from $(\beta \beta )_{2\nu }$ and $(\beta \beta
)_{0\nu }$ decays by the observed shape of the electron spectrum. The
transition rate of Majoron emitting double beta decay together with a
certain degree of confidence in the underlying nuclear physics can be used
to establish constraints on the $G_0$ models in which this decay process (%
\ref{quinze}) can compete with $(\beta \beta )_{0\nu }$ decay\cite{Doi85}.
From Eqs. (\ref{ampeq1}), (\ref{ampeq2}) and (\ref{ampeq3}), 
in the approximation that
the nuclear matrix elements are of the same order of magnitude for all the
processes \cite{Georgi81}, the respective decay rates are related by

\[
\frac{\Gamma _J}{\Gamma _{(\beta \beta )_{0\nu }}}\propto \left( \frac{%
Q\sum\limits_{ij}V_{ei}V_{ej}G_{ij}^R}{\pi \sum\limits_iV_{ei}^2\smallskip\
m_i}\right) ^2R(x)=\frac 1{84}\left[ \frac{Qg_{\text {eff}}}{\pi m_\nu }\right] 
^2R(x),
\]
where  $m_\nu =\left(
\sum\limits_iV_{ei}^2\smallskip\ m_i\right) $ is the mass of the electron
neutrino, $Q$ is the available energy of the decay, and $x\equiv Q/m_e$. The
ratio of phase-space integrals for the processes in question , $R(x)$ is 
\begin{equation}
R(x)=\frac{x^4+14x^3+84x^2+210(x+1)}{x^4+10x^3+40x^2+30(2x+1)}.
\label{dezesete}
\end{equation}

In the Gelmini-Roncadelli model~\cite{Gelm81} the statement $\Gamma _{(\beta
\beta )_{0\nu }}+\Gamma _J\approx 3\Gamma _{(\beta \beta )_{2\nu }}$ from
the tellurium analysis gives the constraint~\cite{Georgi81} 
\begin{equation}
G_{ee}^2\left[ v^2+\frac{m_e^2x^2R(x)}{84\pi ^2}\right] =(30\,\,\mbox{eV})^2
\label{dezoito}
\end{equation}
where $x=1.7$, so that $x^2R(x)=8.4$. Experiments analyzed in terms of the
Majorana mass of the electron neutrino give~\cite{Haxton81} 
\begin{equation}
G_{ee}v<15\,\,\mbox{eV}.  \label{dezenove}
\end{equation}

In the Gelmini-Roncadelli model there are VEV\' s which also contribute 
to the
mass of the $Z^0$ gauge boson. In that case, $Z$ can easily decay into the
Majoron. Measured widths of the $Z$ boson does not have any room for these
particles in this scheme. So, the Gelmini-Roncadelli model have been ruled
out by now. On the other hand in the $G_0$ models only a VEV in the Fermi
scale contributes to the mass of the $Z$ standard boson. To clarify this
mechanism let us consider the 
\[
G_{331}\equiv \mbox{SU(3)}_C\otimes \mbox{SU(3)}_L\otimes \mbox{U(1)}_X
\]
model. In order to generate all masses and to implement the symmetry
breaking hierarchy 
\[
G_{331}\rightarrow G_{321}\rightarrow \mbox{SU(3)}_C\otimes \mbox{U(1)}_{%
\mbox{em}}
\]
we introduce a scalar sector composed of the SU(3) sextet given 
in Eq. (\ref{eq3}) and the SU(3)$_L$ triplets, 
\begin{equation}
\eta \sim ({\bf 1},{\bf 3},0),\quad \rho \sim ({\bf 1},{\bf 3},+1),\quad
\chi \sim ({\bf 1},{\bf 3},-1)  \label{vinte}
\end{equation}
with the vacuum structure 
\begin{equation}
\langle \eta \rangle =(v_\eta ,0,0),\quad \langle \rho \rangle =(0,v_\rho
,0),\quad \langle \chi \rangle =(0,0,v_\chi )  \label{vinteum}
\end{equation}
and 
\begin{eqnarray}
\langle S\rangle =\left( 
\begin{array}{ccc}
v_{\sigma _1} & 0 & 0 \\ 
0 & 0 & v_{\sigma _2} \\ 
0 & v_{\sigma _2} & 0
\end{array}
\right) .  \label{vintedois}
\end{eqnarray}
Notice that even if $v_\eta \approx v_\rho \approx v_{\sigma _1}\approx
v_{\sigma _2}\equiv v_1$ where $v_1$ denotes the usual vacuum expectation
value for the Higgs boson of the standard model, the VEV $v_\chi \equiv v_2$
must be large enough in order to leave the new gauge bosons sufficiently
heavy to keep consistency with low energy phenomenology. In terms of the
adimensional parameters 
\begin{equation}
A\equiv \left( \frac{v_1}{v_2}\right) ^2  \label{vintetres}
\end{equation}
and 
\begin{equation}
t\equiv \frac{g^{\prime }}g  \label{vintequatro}
\end{equation}
where $g$ and $g^{\prime }$ are the SU(3)$_L$ and U(1)$_X$ gauge coupling
constants the mass matrix for the neutral gauge bosons in the $\{W_\mu
^3,W_\mu ^8,B_\mu \}$ basis is 
\begin{equation}
\frac 12M^2=\frac 14g^2\,v_2^2\left( 
\begin{array}{ccc}
3A & \frac 1{\sqrt{3}}A & -2tA \\ 
\frac 1{\sqrt{3}}A & \frac 13(3A+4) & \frac 2{\sqrt{3}}t(A+2) \\ 
-2tA & \frac 2{\sqrt{3}}(A+2) & 4t^2(A+1)
\end{array}
\right)   \label{vintecinco}
\end{equation}
which is a singular matrix due to the vanishing eigenvalue associated to the
photon mass. The nonvanishing eigenvalues, in the limit $A\rightarrow 0$,
are 
\begin{equation}
M_Z^2=\frac 32g^2\frac{1+4t^2}{1+3t^2}v_1^2  \label{vinteseis}
\end{equation}
for the lighter bosons and 
\begin{equation}
M_{Z^{\prime }}^2=\frac 23g^2(1-3t^2)v_2^2  \label{vintesete}
\end{equation}
for the heavier neutral hermitian gauge boson $Z^{\prime }$. On the
other hand the counterparts of charged non-hermitian standard model gauge
boson, have the following mass 
\begin{equation}
M_{W^{\pm }}=\frac 32g^2v_1^2  \label{vinteoito}
\end{equation}
so that in $(331)$ gauge extension 
\begin{equation}
\frac{M_Z^2}{M_{W^{\pm }}^2}=\frac{1+4t^2}{1+3t^2}.  \label{vintenove}
\end{equation}
Comparing with the standard model result, 
\begin{equation}
\frac{M_Z^2}{M_{W^{\pm }}^2}=\frac 1{1-\sin ^2\theta _W},  \label{trinta}
\end{equation}
one obtains 
\begin{equation}
t^2=\frac{\sin ^2\theta _W}{1-4\sin ^2\theta _W}.  \label{trintaum}
\end{equation}
Therefore the theory imposes an upper bound 
\begin{equation}
\sin ^2\theta _W<\frac 14  \label{trintadois}
\end{equation}
with a Landau pole in $\sin ^2\theta _W=1/4$. This constraint on the
electroweak mixing angle $\theta _W$ is a remarkable foresight of the model.

 To sum up, we have discussed the $G_0$ models for $G_W=SU(3)_L$ and $SU(4)_L$   
 and 
 analysed the interaction terms that give rise to Majoron emitting double
 beta decay. 
 Bounds on the effective Majoron neutrino 
 couplings calculated from the reported \cite{Gunt96} half-life limits on 
 Majoron emitting
 decay processes for the nucleus $ ^{76}Ge$ together with nuclear matrix
 elements of Ref. \cite{Hirs96} serve as constraints on the $G_0$ models.
 Increased degrees of freedom allow for the possibility of
 occurence of double beta decay through emission of a charged Majoron($F=-2$) 
 that conserves the more general leptobarion symmetry  
 for the choice $ <\sigma_1^0>=0$. For this kind of decay, the 
 bound on the Majoron-neutrino coupling strength
 is $g_{\text {eff}} < 0.18$. On the other hand a choice of small VEV 
 for the scalar involved that is 
 $<\sigma_1^0>\, \neq 0$ also results in observable majoron emitting 
 double-beta decay. The leptobarion symmetry is broken spontaneously
 in this case. For this process the contribution to $\beta \beta _{\text M}$
 is expected to have many features that are similar to those of the
 Gelmini-Roncadelli model\cite{Gelm81} but no contributions from 
 $\sigma_1^0$ to the invisible width of Z standard boson arise. Only a VEV 
 in  Fermi scale contributes to the mass of Z boson.
 For this process the constraint on the model for the specific theoretical 
 nuclear matrix  element used and the experimental data is
 $g_{\text {eff}} < 2.3X10^{-4} $. 
 A more detailed study of the phenomenology of $\sigma_1^0$ 
 should point out which one is the most natural choice.
 
 By matching the gauge coupling constants at the electroweak scale~\cite{Ng94}
the mass of the new heavy neutral gauge boson, $Z^{\prime }$, is bounded to
be less than 2.2 TeV. An absolute upper limit on the unification scale comes
from $\sin ^2\theta _W<1/4$ ( Eq. (\ref{trintadois})) 
giving $M_{Z^{\prime }}<3.2$ TeV. Unlike most
extensions of the standard model, in which the masses of the new gauge
bosons are not bounded from above, the $G_0$ models would be either realized
or ruled out by the next generation of high energy colliders, or better yet,
at present colliders such as the Tevatron or LEP II. In particular, the $G_W=%
\mbox{SU(3)}_L$ models allow very light scalar bosons, with a neutral one
identified with the standard model Higgs boson~\cite{Tonasse96}. Future
experiments may soon place strong restrictions on these models, thus making
it eminently testable.

\acknowledgements
S. S. S. acknowledges financial support from Conselho Nacional de
Desenvolvimento Cient\'\i fico e Tecnol\'ogico (CNPq) Brazil. 
F.P. was supported by Funda\c c\~ao de Amparo \`a Pesquisa do Estado 
de S\~ao Paulo (FAPESP), Brazil.

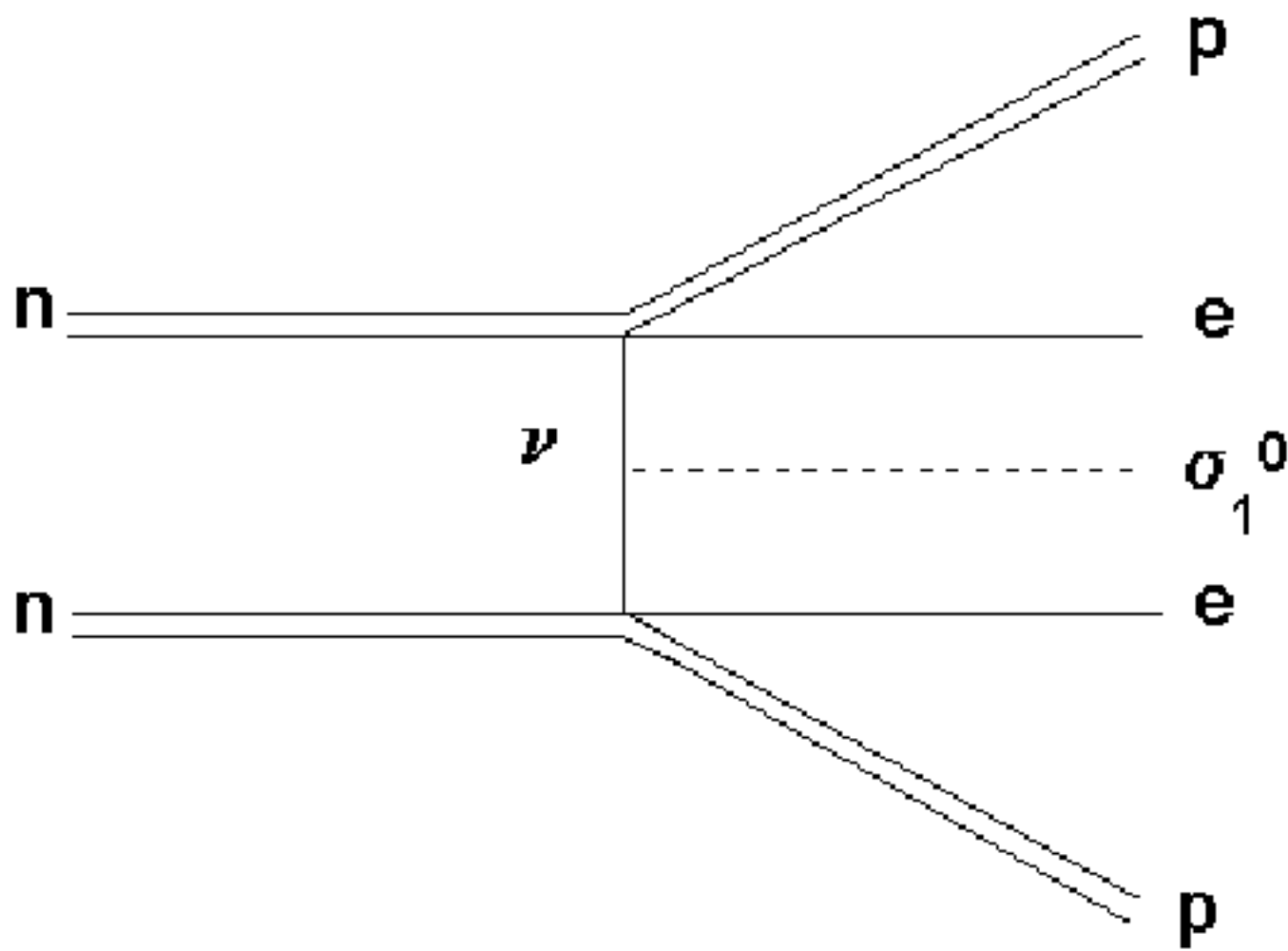
\begin{figure}[tbp]
\caption{Charged Majoron emitting double beta decay, where $\sigma^{1}_0 $ 
carries
 quantum number $F=-2$.}
\end{figure}

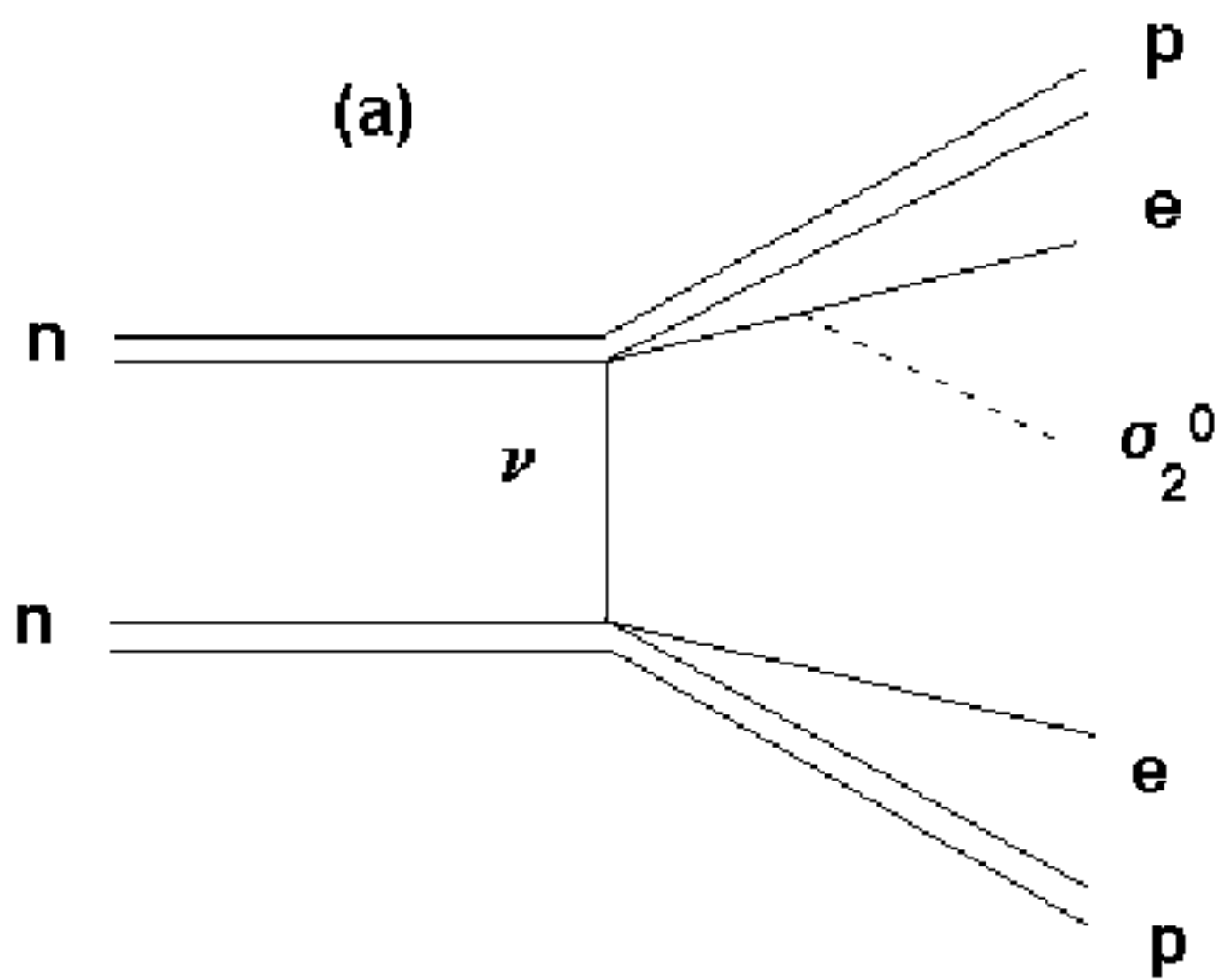
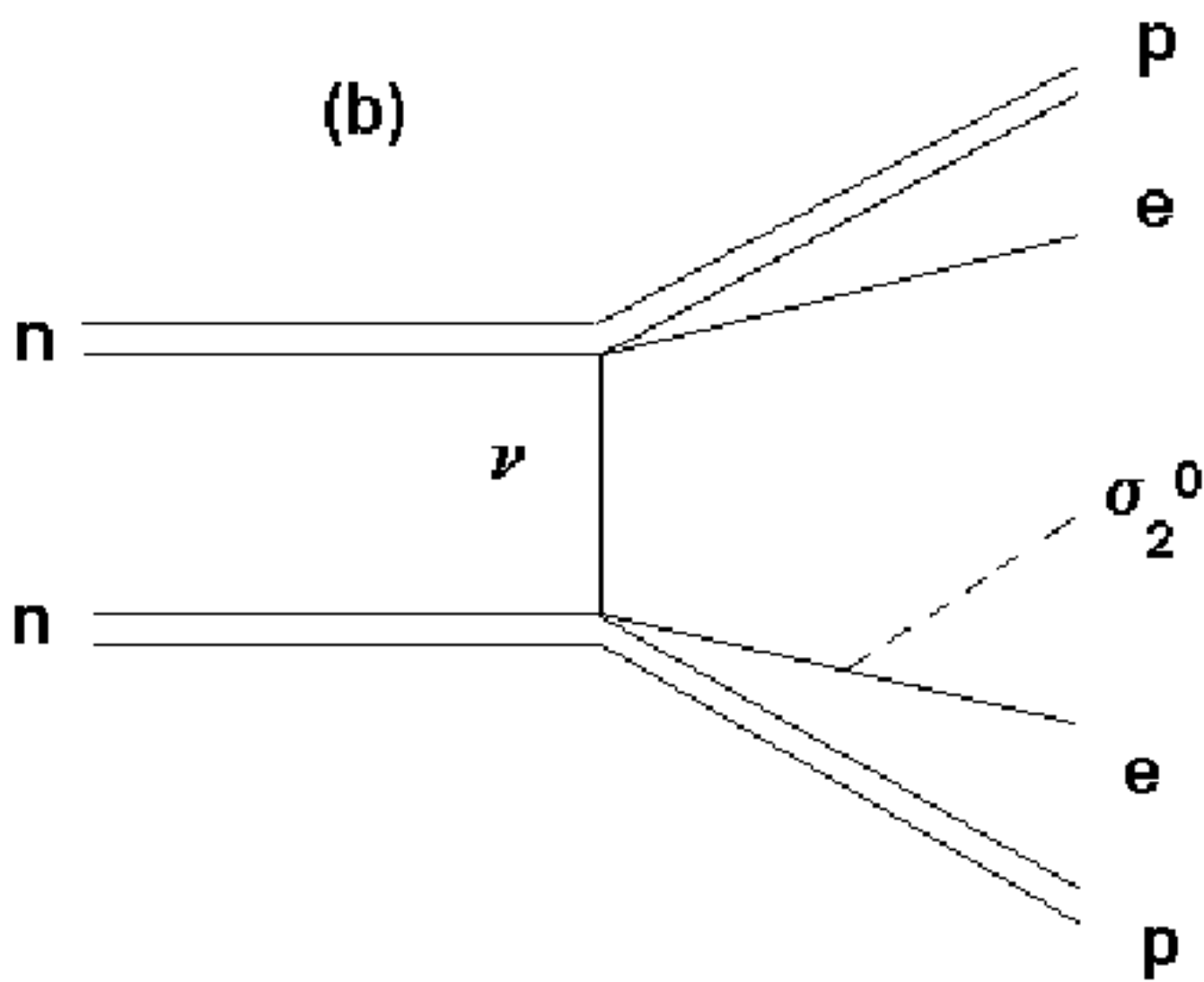
\begin{figure}[tbp]
\caption{Ordinary Majoron emitting double beta
 decay, where $\sigma^{2}_0 $ denotes the ordinary majoron.}
\end{figure}


\begin{references}
\bibitem{Mayer35}  M. Goeppert-Mayer, Phys. Rev. 48, 512 (1935).

\bibitem{Glashow61}  S.L. Glashow, Nucl. Phys. 22, 579 (1961); S. Weinberg,
Phys. Rev. Lett. {\bf 19}, 1264 (1967); A. Salam, in Elementary Particle
Theory, ed. N. Svartholm, Almqvist and Wiksell, 367 (1968); S. L. Glashow,
J. Iliopoulos, and L. Maiani, Phys. Rev. D2, 1285 (1970); M. Kobayashi and
K. Maskawa, Prog. Theor. Phys. 49, 652 (1973).

\bibitem{Elliot87}  S. R. Elliot, A. A. Hahn, and M. K. Moe, Phys. Rev. Lett.%
{\bf \ }59, 2020 (1987).

\bibitem{Furry39}  G. Racah, Nuovo Cim. 14, 322 (1937); W. H. Furry, Phys.
Rev. 56, 1184 (1939).

\bibitem{PDG96}  R. M. Barnett {\it et al}. (Particle Data Group), Phys.
Rev. D54, 1 (1996).

\bibitem{Gunt97}  M. G{\"u}nther, J. Hellmig, G. Heusser, M. Hirsch, 
H.V. Klapdor-Kleingrothaus, B. Maier, H. P{\"a}s, F. Petry, Y. Ramachers,
H. Strecker, M. V{\"o}llinger, A. Balysh, S. T. Belyaev, A. Demehin, A. Gurov, 
I, Kondratenko, D. Kotel\'nikov, V. I. Lebedev and A. M{\"u}ller,  
Phys. Rev. D55, 54(1997).
  

\bibitem{Chik80}  Y. Chikashige, R. N. Mohapatra, and R. D. Peccei, Phys.
Rev. Lett. 45, 1926 (1980); Phys. Lett. B98, 265 (1981).

\bibitem{Gelm81}  G. B. Gelmini and M. Roncadelli, Phys. Lett. B99, 411
(1981).

\bibitem{Voge86}  P. Vogel and M. R. Zirnbauer, Phys. Rev. Lett. 57, 731
(1986).


\bibitem{Doi88}  M. Doi, T. Kotani, and E. Takasugi, Phys. Rev. D37, 2572
(1988).

\bibitem{Geor81}  H. M. Georgi, S. L. Glashow, and S. Nussinov, Nucl. Phys.
B193, 297 (1981).

\bibitem{Burg94}  C. P. Burgess and J. M. Cline, Phys. Rev. D49, 5925 (1994).

\bibitem{Caro93}  C. D. Carone, Phys. Lett. B308, 85 (1993).

\bibitem{Geor74}  H. Georgi and S. L. Glashow, Phys. Rev. Lett. 32, 438
(1974); For a review see P. W. Langacker, Phys. Rep. 72, 185 (1981).

\bibitem{Glashow83}  S. L. Glashow, in {\it The Unity of the Fundamental
Interactions}, Edited by A. Zichichi (Plenum Press, 1983) p.14.

\bibitem{Pisano92}  F. Pisano and V. Pleitez, Phys. Rev. D46, 410 (1992); R.
Foot, O. Hernandez, F. Pisano, and V. Pleitez, Phys. Rev. D47, 4158 (1993).

\bibitem{Fram92}  P. H. Frampton, Phys. Rev. Lett. 69, 2889 (1992).

\bibitem{Mont93}  J. C. Montero, F. Pisano, and V. Pleitez, Phys. Rev. D47,
2918 (1993).

\bibitem{Volo88}  M. B. Voloshin, Sov. J. Nucl. Phys. 48, 512 (1988).

\bibitem{Foot94}  R. Foot, H. N. Long, and T. A. Tran, Phys. Rev. D50, R34
(1994); F. Pisano and V. Pleitez, Phys. Rev. D51, 3865 (1995).

\bibitem{Pisano96}  F. Pisano, Mod. Phys. Lett. A11, 2639 (1996)

\bibitem{Ozer94}  M. \"Ozer, Phys. Lett. B337, 324 (1994); F. Pisano, J. A.
Silva-Sobrinho, and M. D. Tonasse, Phys. Lett. B388, 338 (1996).

\bibitem{Ng94}  D. Ng, Phys. Rev. D49, 4805 (1994).

\bibitem{Barb89}  R. Barbieri and R. N. Mohapatra, Phys. Lett. B218, 225
(1989); J. Liu, Phys. Lett. B225, 148 (1989).

\bibitem{Babu88}  K. S. Babu, Phys. Lett. B203, 132 (1988).

\bibitem{Plei93}  V. Pleitez and M. D. Tonasse, Phys. Rev. D48, 5274 (1993).


\bibitem{Krmp94}  F. Krmpotic and S. Shelly Sharma, Nucl. Phys. A572, 329
(1994).

\bibitem{Caur96}E. Caurier, F. Nowaski, A. Poves and J. Retamosa, 
 e-print Archive nucl-th/9601017.
 
\bibitem{Barb96}  C. Barbero, J. M. Cline, F. Krmpotic, and D. Tadic, Phys.
Lett. B371, 78 (1996).

\bibitem{Hirs90}  J. Hirsch and F. Krmpotic, Phys. Rev. C 41, 792 (1990); J.
Hirsch and F. Krmpotic, Phys. Lett. B 246, 5 (1990).

\bibitem{Hirs96}M.Hirsch, H. V. Klapdor-Kleingrothaus, S. G. Kovalenko 
 and H. P\"s, Phys. Lett. B 372, 8 (1996): J. Hellmig et al., in
 { Proceedings of the International Workshop on Double Beta Decay
 and Related Topics}, edited by H. V. Klapdor-Kleingrothaus and
 S. Stoica (World Scientific, Singapore, 1996).


\bibitem{Gunt96}  M. G{\"u}nther, J. Hellmig, G. Heusser, M. Hirsch, 
H.V. Klapdor-Kleingrothaus, B. Maier, H. P{\"a}s, F. Petry, Y. Ramachers,
H. Strecker, M. V{\"o}llinger, A. Balysh, S. T. Belyaev, A. Demehin, A. Gurov, 
I, Kondratenko, D. Kotel\'nikov, V. I. Lebedev and A. M{\"u}ller,  
 Phys. Rev. D54, 3641(1996).


\bibitem{Doi85}  M. Doi, T. Kotani and E. Takasugi, Progr. Theor. Phys.
Suppl. 83, 1 (1985).

\bibitem{Georgi81}  H. M. Georgi, S. L. Glashow and S. Nussinov, Nucl. Phys.
B193, 297 (1981).

\bibitem{Haxton81}  W. C. Haxton, G. J. Stephenson, Jr and D. Strottman,
Phys. Rev. Lett. 46, 698 (1981).

\bibitem{Tonasse96}  M. D. Tonasse, Phys. Lett. B381, 191 (1996)
\end{references}
\end{document}